\documentclass[aps,prd,reprint,groupedaddress, showpacs, showkeys]{revtex4-1}
\usepackage{graphicx}
\usepackage{txfonts}

\begin{document}


\title{Cosmic Dark Energy Emerging from Gravitationally Effective Vacuum Fluctuations}


\author{Bruno M. Deiss}
\email[]{deiss@em.uni-frankfurt.de}
\affiliation{Institute for Theoretical Physics (ITP), Goethe-University,
              Max-von-Laue-Stra\ss e 1, D-60438 Frankfurt am Main, Germany}



\begin{abstract}
Astronomical observations indicate an accelerated cosmic expansion, the cause of which is explained by the action of `dark energy'. Here we show that in discrete expanding space-time,
only a tiny fraction of the vacuum fluctuations can become gravitationally effective and act as a driving `dark' agent. The analytically derived effective vacuum energy density is found to be closely related to the critical cosmic energy density, thus helping to solve the cosmological constant problem as well as the coincidence problem. The proposed model implies that in the present day universe only the vacuum field of the photon and that of the lightest neutrino contribute to the effective vacuum. This allows one to fix the neutrino masses within a narrow range. The model also implies that the (real) universe has to be considered as a thermodynamically open system which exchanges energy and momentum with the virtual reservoir of the vacuum.
\end{abstract}

\pacs{95.36.+x, 98.80.-k, 04.60.-m, 14.60.Pq}
\keywords{dark energy theory, modified gravity, quantum cosmology}

\maketitle

%
\section{Introduction}
\label{sec:introduction}

Over the past decade, astronomical studies have converged toward a cosmic expansion history that involves a recent accelerated expansion (e.g. \cite{fth2008}, \cite{blanchard2010}, \cite{weinberg2012}). A widely accepted model is that of a dominant cosmic energy component (`dark energy') with negative pressure, which provides the dynamic mechanism for the acceleration. 

In the present day universe dark energy accounts for nearly three quarters of the whole cosmic energy budget. Yet, compared to quantum field theoretical expectations, its density is about 120 orders of magnitude smaller, a challenge occasionally dubbed the (old) `cosmological constant problem' \cite{weinberg1989}.  Another challenge is the `cosmological coincidence problem': obviously, we happen to live in a cosmic era where the densities of the matter and dark energy components are of the same order of magnitude, although one expects these quantities to behave very differently in the course of the cosmic evolution. 

From a conceptual point of view, the simplest solution to the cosmic acceleration problem consists in the addition of a cosmological constant $\Lambda$ to Einstein's equations. This is formally equivalent to the introduction of a constant dark energy density $\rho_{\mathrm{DE}} = \rho_{\Lambda}= \Lambda c^4/(8\pi G)$ with negative pressure. At present, the astronomical observations actually favour the `$\Lambda$CDM-model' where one adopts, besides cold dark matter (CDM), a cosmological constant. 
The fraction $\Omega_{\Lambda, 0} = \rho_{\Lambda}/\rho_{\mathrm{crit, 0}}$, where $\rho_{\mathrm{crit, 0}} = 3c^2H^2_0/(8\pi G)$ denotes the local critical cosmic energy density, has been recently evaluated to be of 
\cite{komatsu2011}
   \begin{equation}
				\Omega_{\Lambda, 0} = 0.725 \pm 0.016
				.
		\label{OmegaLambda}
		\end{equation}

Dispite the $\Lambda$CDM-model's practical success, from a theoretical point of view it is somewhat unsatisfying to just introduce a new fundamental constant. Hence, a large number of theoretical concepts and physical models has been proposed in order to explain the origin and evolution of the dark energy component 
[e.g. \cite{calkam2009}, \cite{ruiz2010}, and references therein]. 
Yet, the proposed models often suffer from the need of fine-tuning of otherwise unexplained free parameters. 

Vacuum energy would be the most natural explanation for such a constant energy contribution and it is actually one of the prime candidates for the solution of the dark energy problem. However, besides the above mentioned cosmological constant problem, it is even unclear today how the vacuum could gravitationally interact with space-time.
It is the aim of the present paper to introduce a new perspective on this issue, which allows to construct a viable dark energy model. 

The paper is organized as follows: In section \ref{sec:effective_vacuum} we propose a new gravitationally effective vacuum model, expound the underlying concepts and derive an analytic expression of the expected amplitude of the effective vacuum energy density. In section \ref{sec:local universe} we firstly estimate the resulting amplitude in the local universe. The inferred evolution of the cosmic expansion rate is then dealt with in section \ref{sec:cosmic evolution}, where we assume a spatially homogeneous universe. In section \ref{sec:Econserv} we consider the problem of energy conservation and the equation-of-state parameter of the effective vacuum component. In section \ref{sec:inhomogeneity} we briefly discuss how spatial inhomogeneities would affect the amount of the effective energy density.
Finally, section \ref{sec:conclusions} contains our conclusions. 

%
\section{Gravitationally effective vacuum energy}
\label{sec:effective_vacuum}

General relativity (GR) postulates that gravitation couples universally to all forms of energy and momentum. Hence, in continuous space-time vacuum fluctuations of all length-scales should gravitate; however, the respective total energy density would then diverge toward unphysical values. A widely accepted `solution' to this problem is the assumption that, due to some unspecified symmetry, fluctuations of the vacuum fields actually do not couple to gravitation. Although violating the general relativistic postulate, this appears to be justified in view of the successful renormalization procedures performed in quantum field theory, where the energy density of the vacuum can be reset to zero.

Here we propose a different perspective, showing that vacuum fluctuations actually do gravitate; but that this happens to be the case essentially only for fluctuations within a naturally given limited range of energies, thus meeting the above mentioned postulate of general relativity. However, this is achieved at the cost of violating another GR-postulate, that of \emph{continuous} space-time. 

It appears that the cosmological constant problem emerges in some sense from a one-directional perspective: one only considers the \emph{action} of the total sum of the vacuum fluctuations on the general relativistic space-time. Instead, we assert that it is essential to consider a \emph{back-reaction}, as well; however, not on the total sum of the vacuum fluctuations, which would make no sense, but on each individual virtual particle. 

To be more specific, we assume that virtual fluctuations, just as real photons, may be subjected to the overall expansion of space during their limited lifetime. In continous space-time this assumption alone does not yet suffice to solve the problem. Below we show that in \emph{process-related discrete} space-time,
only particles within a limited energy range actually \emph{interact}. Hence, in the absence of interaction, as it is in the case for virtual particles especially at energies of the highest level, there is basically no gravitational coupling. 

In other words, the principle adopted here is the following: gravity implies interaction; virtual fluctuations that do not interact remain literally virtual and decouple from gravity, while fluctuations that interact become "real" in the sense that their energies contribute to what we refer to as the \emph{effective} vacuum energy density. In the specific case of expanding space-time this becomes noticeable as cosmic dark energy.
In order not to obscure the underlying concept, we focus in what follows on basic arguments.

\subsection{Vacuum fluctuations}
\label{sec:vacuum fluctuation}

The vacuum ground state of a field $i$ (e.g. photons, neutrinos, etc.) can be described by an ensemble of free harmonic oscillators of angular frequency $\omega$ and wavevector $\vec{k}$, each with a zero-point energy of 
   \begin{equation}
      E_{i}(\omega)
      =  \hbar \omega_{i}/{2} .
		\label{epsdef}
		\end{equation}

Summing the zero-point energies of all modes of all fields, the total vacuum energy density is given by
   \begin{equation}
   		\rho_{\mathrm{vac}}  =  \sum_{i} \rho_{i} =
   		\sum_{i} \frac{g_{i}}{(2\pi)^{3}}
   		\int_{k_{0,\, i}}^{k_{1,\, i}}
   		(\hbar \omega_{i} /2)
   		\:\mathrm{d}^{3}\vec{k}
   		\label{rhovac} .
   \end{equation}   	
In the following, we adopt the dispersion relation
  \begin{equation}
      \hbar \omega_{i}
       =  \sqrt{ m_{i}^2 c^4 + p^2 c^2}
      =  \sqrt{ \left(m_{i} c^2\right)^2 + \left(\hbar c k \right)^2},
		\label{omegadef}
		\end{equation}	
where $m_{i}$, $p$ and $\hbar = h/2\pi$ denote respectively the particle mass of species $i$, the momentum and the reduced Planck constant.	In each term of the sum (\ref{rhovac})	
the factor $g_{i}$ accounts for the degrees of freedom (e.g. helicity) as well as the phase factor of the specific field under consideration. Hence, $g_{i}$ is negative for fermionic fields, while positive for bosonic ones. 

As is well-known, each integral in (\ref{rhovac}) diverges towards an unphysically large value unless the upper limit of integration is considerably reduced. The latter actually happens to be the case in process-related discrete space-time as is shown below; one then arrives at the \emph{effective} vacuum energy density. 

\subsection{Interaction limit}
\label{sec:interaction limit}

The notion of \emph{discrete} space-time is usually associated with the assumption that all physical length-scales are bound from below by a presumed minimal length (`Planck limit'). Instead of this, we assert that the assumed discreteness primarily involves a lower limit of the \emph{changes} of length-scales, $\Delta L$, and of time-scales, $\Delta T$, in the course of an interaction; the usual Planck limit is then an implicit corollary. Though this assertion appears to be self-evident, it has not, to the knowledge of the author, been explicitely considered in the literature so far. However, as is shown below, this seemingly minor modification establishes a conceptional shift that turns out to play a key role in order to tackle the well-known UV divergency problem. Hence, the \emph{scale differences} $\Delta L$ and $\Delta T$ must fulfill the conditions
 		\begin{equation}
   		   	|\Delta L| \geq  \eta l_{\mathrm{Pl}} \ \ \ \ \ \mathrm{and}\  \ \ \ |\Delta T| \geq  \eta l_{\mathrm{Pl}}/c ,
   \label{DeltaL}             	
   \end{equation}
where $l_{\mathrm{Pl}}$ denotes the Planck length. The latter is related to the gravitational constant $G$ by
   \begin{equation}
   		   	l_{\mathrm{Pl}} = \left(\hbar G/c^3\right)^{\frac{1}{2}} = 1.616 \times 10^{-35} \mathrm{m} .
   \label{lPlanckdef}                        	
   \end{equation}   
The factor $\eta$ serves as a structure parameter which allows one to account for the possibility that the effective minimal length scale may be a multiple (e.g. $\eta = \sqrt{8\pi}$) of the Planck length. 

The total energy change, $\Delta E_{i}$, that a virtual particle of energy $E_{i}$ experiences in the course of an interaction (in the present paper: due to the expansion of space-time considered in the reference frame where the cosmic microwave background is isotropic) is given by
   \begin{equation}
				|\Delta E_{i}| = \frac{\hbar}{2} \left|\Delta \omega_{i} \right|
									 = \hbar \pi \left|\Delta \frac{1}{T_{i}} \right|
									 = \hbar \pi \frac{\left|\Delta T_{i}\right|}{T_{i}^2} 
									 = \frac{E_{i}^2}{\pi \hbar} \left|\Delta T_{i} \right| ,
				\label{DeltaEdef}
   \end{equation}
where $\Delta T_{i}$ denotes the respective change in period. From condition (\ref{DeltaL}) and expression (\ref{DeltaEdef}) one obtains the \emph{interaction limit} (or, \emph{process-related coupling scale})
   		\begin{equation}
   		   	E_{i}^2  \leq  \pi \hbar c |\Delta E_{i}| / (\eta l_{\mathrm{Pl}} )  
   		    = (\pi/\eta) E_{\mathrm{Pl}} |\Delta E_{i}| \:
	,
	 \label{E2leq}  	
   \end{equation}
where $E_{\mathrm{Pl}} = \hbar c /l_{\mathrm{Pl}}$ denotes the Planck energy. Hence, only virtual particles with energies below a given limit actually take part in the interaction. More energetic particles violate the constraints (\ref{DeltaL}) and basically do not interact. It is widely believed that the presumed microscopic granularity of space-time becomes important only when the physical scales approach the Planck scale. From (\ref{E2leq}),
however, we conclude that the underlying process-related discrete structure actually affects macroscopic scales as well. And this happens to be the case especially in \emph{low-energy} interactions where $|\Delta E_{i}|$ is small. 

Note that constraints (\ref{DeltaL}) and (\ref{E2leq}) imply a concept of the discret structure of space-time, which is quite different to its common notion. Whilst the latter is generally based on the assumption that space-time \emph{is} granular at the Planck-scale (in the sense of Wheeler's "quantum foam" \cite{wheeler1962}), the proposed concept relies on the assumption that the microstructure is such that it only constrains physical \emph{processes}. In a sense, the new notion represents a conceptual shift from \emph{being} to \emph{becoming}. Thus, the coupling scale (\ref{E2leq}) is basically and essentially process-related and the respective cutoff depends on the kind of interaction under consideration.

\subsection{Expanding space time}
\label{sec:expanding space time}

In the following we apply condition (\ref{E2leq}) to the vacuum fluctuations in a large-scale homogeneous and isotropic universe which expands at a rate $H = (\mathrm{d} a /{\mathrm{d} t})/ a$, where $a$ represents the scale-length of the universe. In analogy to the cosmological redshift of photons, we assume that the spatial expansion also affects the energy and momentum of virtual particles within their limited lifetimes. 

According to the assumptions (\ref{DeltaL}), the expansion of space may also be considered as a quantized process, where on microscopic scales discrete space elements pop up stochastically at a specific rate. Here we confine ourselves to effects of lowest order. To this end, we consider $H$ as a \emph{mean} rate averaged over appropriate length and time scales. In a sense, this procedure is equivalent to that in thermodynamics, where macroscopic quantities emerge from coarse-graining of the atomic scales. In the same line of reasoning, and for the sake of simplicity, we consider only \emph{mean} lifetimes, $\tau_{i}$, of each species of the virtual particles. 

In expanding space-time, the wavelength of each particle, real or virtual, becomes redshifted at a rate $\dot{\lambda} / \lambda = H$. Within a short time interval, $\tau$, this amounts to $|\Delta \lambda| \approx \lambda H \tau$. The corresponding momentum change 
is given by 
  \begin{equation}
   		   	  |\Delta p| = \hbar |\Delta k| = \hbar \, 2\pi \,|\Delta (1/\lambda)| = \hbar \, 2\pi \,|\Delta \lambda| /\lambda^2 \approx p H \tau \ .
   \label{Deltap}               	
   \end{equation}
Hence, the total momentum change a virtual particle experiences due to the expansion of space within its (energy-dependent) lifetime amounts to  
   \begin{equation}
   		   	|\Delta p_{i}| = 
   		   	p H \tau_{i} = p H \hbar/(2E_{i}),
   \label{Deltaptotal}               	
   \end{equation}
where the latter relation follows from Heisenberg's uncertainty principle. Considering relations (\ref{epsdef}), (\ref{omegadef}) and (\ref{Deltaptotal}) the respective total energy change is given by
  \begin{equation}
   		   	|\Delta E_{i}| = H \hbar \ p^2 c^2/(8E^2_{i}).
   \label{DeltaEtotal}               	
   \end{equation}

Note that expression (\ref{DeltaEtotal}) is only meaningful for those particles that meet condition (\ref{E2leq}).   
From (\ref{E2leq}) and (\ref{DeltaEtotal}) we find that
		\begin{equation}
   		   	E_{i}^2 \leq  
   		   	(p c/2) \sqrt{\pi E_{\mathrm{Pl}} \hbar H /(2 \eta)} \ \equiv \ 
   		   	(p c/2) \; D(H)
   		   	,             				
   \label{Epcond}  	
   \end{equation} 
where $D(H)$ defines a new characteristic cosmic energy scale that depends on the expansion rate $H$. 

In the local universe the characteristic `interaction scale' $D$ in (\ref{Epcond}) amounts to
		\begin{equation}
   		   D_0 \equiv D(H_0)
   		   = \sqrt{h_{100}/\eta} \left(6.40 \times 10^{-3} \rm{eV}\right)
    		   	 ,       				
   \label{DH0}  	
   \end{equation}
where $h_{100} = H_0/(100 \mathrm{km/s/Mpc})$ denotes the normalized local Hubble parameter.
Its observationally determined value is in the range of 
[e.g \cite{komatsu2011}, \cite{riess2011}, \cite{beutler2011}] 
		\begin{equation}
   		   h_{100} = 0.64 - 0.76    				
   \label{h100def}  	
   \end{equation}
Note that the amount of $D_0$ (\ref{DH0}) is of the order of the observed dark energy scale as well as of the conjectured neutrino mass scale 
[e.g. \cite{gandoetal2010}, \cite{barry2011}]. 
We emphasize that, as an important outcome of the proposed model, this `macroscopic' cosmic energy scale naturally emerges from the presumed process-related microscopic structure of space-time. 

This characteristic scale is not a constant but evolves in the course of the cosmic history as
		\begin{equation}
   		   D(H) = D_0 \sqrt{H/H_0}   		 
   		      	 .       				
   \label{DHevolv}  	
   \end{equation}

Obviously, only wavenumbers of a finite range satisfy condition (\ref{Epcond}). Employing (\ref{epsdef}) and (\ref{omegadef}), the respective mass-dependent limits entering the integrals in the sum (\ref{rhovac}) are given by
		\begin{equation}
          	k_{1/0,\: i}(H) =  D(H)/(\hbar c) \times \left[1 \pm \sqrt{1-x_{i}^2 (H)}\:\right],
   \label{kiminmax} 
		\end{equation}
where $x_{i}$ denotes the relative mass defined by
\begin{equation}
				x_{i} (H) = m_{i} \: c^2/D(H) < 1.
    \label{xdef} 
		\end{equation}
		
Expression (\ref{kiminmax}) exhibits an interesting relationship between the UV cutoff $k_1$ and an apparent IR cutoff $L_H$, defined by $L_H \propto 1/H$, via the interaction scale $D(H) \propto \sqrt{E_{\mathrm{Pl}}/L_H}$. Deduced from more general reasons, a similar functional relationship that should hold in effective field theories has been proposed by Cohen et al. \cite{cohkanelson1999}. And it is one of the basic assumptions of the so-called holographic dark energy models discussed in the literature [e.g. \cite{xu2012}, \cite{duran2012}]. The latter suggests a close but still to be examined relationship between the effective vacuum model proposed in the present paper and the holographic principle \cite{tHooft2009}. In some sense, the new notion of a process-related coupling scale (\ref{E2leq}) provides a physical explanation for the holographic principle. 

Applying the wavenumber limits (\ref{kiminmax}) to the integral (\ref{rhovac}) and substituting $D$ (\ref{Epcond}) we obtain
   \begin{equation}
   		\rho_{\mathrm{vac}}(H) = 
   		\frac{D^4(H) }{\pi^2\hbar^3 c^3}\: W(H) 	
   		=			\frac{\hbar H^2 W(H)}{4c \eta^2 l_{\mathrm{Pl}}^2}    
   		    =    \frac{c^2 H^2 W(H)}{4 \eta^2 G} \ ,	 
   		\label{rhovacW} 
   \end{equation}   
where the dimensionless quantity $W(H)$ is defined by the sum
   \begin{equation}
   		W(H)	=   \sum_{i} g_{i} \: f_{i}(H) .
   		\label{Wdef} 
   \end{equation}  
Each function
$f_{i}(H) \equiv f(x_i (H))$ can be regarded as a weighting function that, for a given expansion rate $H$, specifies the individual contribution of each field $i$ with relative particle mass $x_{i}$ (see Appendix). 
Figure \ref{fig:fi} displays $f_{i}$ as a function of $x_{i}$, showing that the contribution of a vacuum field monotonically decreases with increasing particle mass. The most important finding is that virtual particles with relative mass $x_{i} \geq 1$, or, $m_{i} \geq D(H)/c^2$, actually do not contribute to the effective vacuum, i.e., they decouple from expanding space-time. 

		\graphicspath{{d:/research/TeX/PhysRev/PRD/publ/}}
		\begin{figure}
				\includegraphics[width=\linewidth]{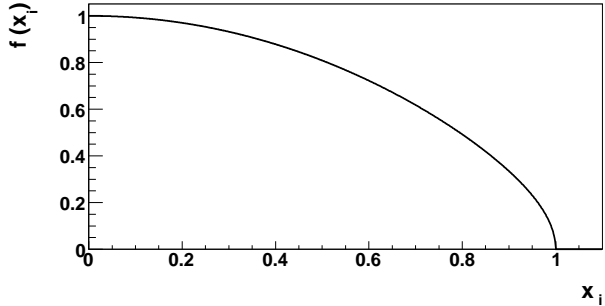}
		 		\caption{Specific weighting factor $f(x_i)$ (\ref{intgeneral}) as a function of relative particle mass $x_{i}$ (\ref{xdef}). Massive fields where $x_{i} \geq 1$ do not contribute to the effective vacuum.}
		\label{fig:fi}
		\end{figure}

Expression (\ref{rhovacW}) can be recast to
   \begin{equation}
   		\rho_{\mathrm{vac}}(H)  
   				   		    =		 \frac{2\pi}{3 \eta^2} \: W(H)\: \rho_{\mathrm{crit}}(H) \ ,
   		\label{rhovactotcrit} 
   \end{equation}  
or,
   \begin{equation}
   		\Omega_{\mathrm{vac}}(H)  
   									\equiv \rho_{\mathrm{vac}}(H) /\rho_{\mathrm{crit}}(H) 
   				   		    =		 \frac{2\pi}{3 \eta^2} \: W(H) \ ,
   		\label{omegavac} 
   \end{equation}     
showing the close relation to the cosmic critical energy density. For a given expansion rate $H$ the function $W(H)$ (\ref{rhovacW}) is a finite sum over weighted numbers $g_i$ whose absolute values are each of order unity. Thus one expects  that the effective vacuum energy density and the critical energy density are both of the same order of magnitude. Note that the total weighting function $W(H)$, therefore also $\rho_{\mathrm{vac}} (H)$, can alter their signs in the course of the evolution of the expansion rate $H$. 

Identifying $\rho_{\mathrm{vac}}(H)$ in (\ref{rhovactotcrit}) with the cosmic dark energy density $\rho_{\mathrm{DE}}$ provides an interesting constraint between the evolution of the cosmological quantities $\rho_{\mathrm{DE}}$ and $H$, the structure parameter $\eta$ and the elementary particle parameters $g_{i}$ and $m_{i}$. Vice versa, this constraint may serve as a critical test for the proposed model.
In the following sections we demonstrate that a consistent and surprisingly simple model can be constructed. 

\section{Local universe}
\label{sec:local universe}
 
As a first crucial test for the reliability of the proposed model we evaluate the resulting present day effective vacuum energy density 
  \begin{equation}
   		\Omega_{\mathrm{vac,0}}
   									=		 \frac{2\pi}{3 \eta^2} \: W(H_0) 
   									\equiv \frac{2\pi}{3 \eta^2} \: W_0.
   		\label{omegavac0} 
   \end{equation}     
   
Assuming that the structure parameter $\eta$ (\ref{DeltaL}) is of order unity, the characteristic local interaction scale $D_0$ (\ref{DH0}) is within the range of the neutrino mass scale, as noted above. Other hypothetical elementary particles like axions or sterile neutrinos with masses of the order of $D_0$ might exist. Here, however, we confine ourselves to considering only free elementary particles that are so far experimentally verified. Within the range of $D_0$ these are comprised of photons and the three neutrino flavours and mass eigenstates. We also disregard massless gluon fields as these vacuum fields are relevant presumably only within hadrons. 

The experimentally determined larger (atmospheric) and smaller (solar) mass-squared differences of the neutrino masses are respectively $\left|\Delta m_{A}^2\right| = (2.43 \pm 0.13) \times 10^{-3} (\mathrm{eV}/c^2)^2$ and $\Delta m_{S}^2 = 7.50 \:(-0.20/+0.19) \times 10^{-5} (\mathrm{eV}/c^2)^2$ [e.g. \cite{bilenky2012}, and references therein]. Compared with (\ref{DH0}), one finds $\left|\Delta m_{A}^2\right| \gg (D_0/c^2)^2$ and even $\Delta m_{S}^2 > (D_0/c^2)^2$. In view of the limit (\ref{xdef}) this implies that besides photons only the vacuum field of the lightest neutrino can, if at all, contribute to the present day dark energy density, irrespective of the neutrino mass hierarchy. Hence, the local value of the total weighting function is given by
\begin{equation}
   		  W_0 = g_{\mathrm{ph}} - \left|g_{\nu}\right|	f_{\nu, \,\mathrm{l},\, 0}				
					, 
   		  \label{W0}                	
   \end{equation}
where respectively $g_{\mathrm{ph}}$ and $g_{\nu}$ refer to the photon and neutrino field and the index `l' stands for the lightest neutrino species.

To proceed, we adopt $g_{\mathrm{ph}} = 2$ according to the two possible directions of photon polarization. The fermionic neutrinos could in principle exhibit four degrees of freedom (Dirac particles). This would imply a lower bound of the mass of the lightest neutrino of $m_{\nu, \mathrm{l}} \gtrsim D_0/c^2$, since $f_{\nu, \, \mathrm{l}, \, 0}$ in (\ref{W0}) must then be very small or even vanish. However, experimental findings indicate that neutrinos and anti-neutrinos each come along with only one spin helicity (mutually oppositely orientated), suggesting that neutrinos are Majorana particles. This shows that, in principle, the proposed model can offer valuable clues to this issue. In the present paper, however, we refrain from further consideration of this important question of neutrino physics but simply refer to the current observations which imply $g_{\nu} = - 2$ in eq. (\ref{W0}). Adopting this, the local $\Omega$-parameter of dark energy finally becomes
    \begin{equation}
   		  \Omega_{\mathrm{DE,0}} \equiv
   		  \Omega_{\mathrm{vac,0}} = 4\pi/(3 \eta^2) \times \left( 1 - f_{\nu, \, \mathrm{l}, \, 0}	\right)					
					.
   		  \label{OmegaReduc2}                	
   \end{equation}
Hence, two free parameters are left to adjust (\ref{OmegaReduc2}) to the observations: the structure parameter, $\eta$, and the relative mass of the lightest neutrino, $x_{\nu, \, \mathrm{l}, \, 0} = m_{\nu, \mathrm{l}} \, c^2 /D_0$, which determines the value of the specific weighting factor 
$f_{\nu, \, \mathrm{l}, \, 0}$. 

Within the frame of the adopted simplifications, from (\ref{OmegaReduc2}) one can draw a first interesting conclusion: even the lightest neutrino mass must be non-zero, otherwise $f_{\nu, \, \mathrm{l}, \, 0} = 1$ and $\Omega_{\mathrm{DE,0}}$ would vanish. 

If on the other hand the lightest neutrino were more massive than $D_0/c^2$ (\ref{DH0}), 
$f_{\nu, \, \mathrm{l}, \, 0}$ would vanish. One then infers a lower limit for the structure parameter of $\eta \geq \sqrt{4\pi/(3)} \approx 2.4$, as $\Omega_{\mathrm{vac,0}}$ should not exceed unity. In the intermediate case, i.e., $0 < m_{\nu, \mathrm{l}} < D_0/c^2$, the structure parameter $\eta$ can also be smaller than 2.4. Vice versa, assuming a certain value of $\eta$ constrains the mass of the lightest neutrino from above.    

This allows, at least in principle, to use the proposed effective vacuum model as a phenomenological test for the structure parameter $\eta$. However, since we do not aim to perform extended parameter studies in the present paper, we simply adopt $\eta = 1$ in what follows. That means, we assume that the Planck length itself represents the presumed effective process-related minimal length-scale of space-time. 


		\graphicspath{{d:/research/TeX/PhysRev/PRD/publ/}}
		\begin{figure}
				\includegraphics[width=\linewidth]{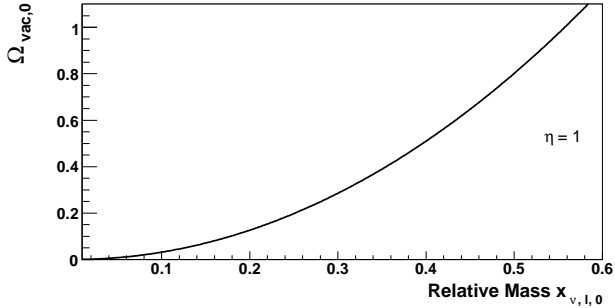}
		 		\caption{The local effective vacuum energy density parameter $\Omega_{\mathrm{vac, 0}}$ (\ref{OmegaReduc2}) as a function of the relative mass of the lightest neutrino $x_{\nu, \mathrm{l, 0}} = m_{\nu, \mathrm{l}} \, c^2 /D_0$. The structure parameter (\ref{DeltaL}) is chosen to be $\eta = 1$. 
		 		}
				\label{fig:Omega}
		\end{figure}

Figure \ref{fig:Omega} displays the amount of the local effective vacuum energy density parameter, $\Omega_{\mathrm{vac,0}}$, as a function of the relative mass of the lightest neutrino $x_{\nu, \, \mathrm{l}, \, 0}$. If $\Omega_{\mathrm{vac,0}}$ is taken to be identical to the observationally determined dark energy parameter, $\Omega_{\mathrm{DE,0}}$, a reliable amplitude range is given by $0.5 \leq \Omega_{\mathrm{vac,0}} \leq 1$. This in turn confines the relative neutrino mass to be $0.40 \leq x_{\nu, \mathrm{l, 0}} \leq 0.56$. Considering the observationally given span of the Hubble parameter (\ref{h100def}), one then infers that the amount of the lightest neutrino mass is constrained by the rather narrow interval $2.3 \lesssim m_{\nu, \mathrm{l}} \lesssim 3.2 \; \mathrm{meV}/c^2$.

In order to estimate an upper limit of the respective summed neutrino mass $M_{\nu} = m_{\nu, \mathrm{l}} + m_{\nu, 2} + m_{\nu, 3}$, we adopt 
$m_{\nu, \mathrm{l}} = 3.2 \; \mathrm{meV}/c^2$ and an inverted mass ordering. From this and the measured mass-squared differences, we find $M_{\nu} \lesssim 0.11\; \mathrm{eV}/c^2$, a value well below current upper bounds determined by neutrino oscillation experiments and by observational cosmology [e.g. \cite{bilenky2012}, \cite{gonzales2010}, \cite{wong2011}, \cite{jose2011}]. This shows that the proposed gravitationally effective vacuum concept provides a consistent model to resolve the (old) cosmological constant problem.

In addition, the model predicts that the mass of the heaviest neutrino is $0.05 \; \mathrm{meV}/c^2$ at the most. Hence, the KATRIN experiment \cite{wolf2010} with its sensitivity of $0.2 \; \mathrm{meV}/c^2$ may serve as a crucial experiment for the proposed model.

\section{Cosmic evolution}
\label{sec:cosmic evolution}	

The amplitude of the derived effective vacuum energy density parameter $\Omega_{\mathrm{vac}}(H)$ (\ref{omegavac}) depends on the value of the cosmic expansion rate $H$ which itself affects the cosmic evolution. At present, the astronomical data point to a cosmic history that can be reliably represented by a simple $\Lambda$CDM-model \cite{komatsu2011}. Therefore, in the following we choose the $\Lambda$CDM-model as a reference. 

To this end, we confine ourselves to considering a homogeneous, isotropic and spatially flat universe whose total cosmic energy budget comprises of only matter and dark energy. Depending on the considered model, the latter is represented either by a cosmological constant or by the effective vacuum energy.

The density of the matter component, assumed to be pressureless and energetically isolated from the dark energy component, scales with redshift $z = (a_0-a)/a$ as $\rho_{\mathrm{m}}(z)/\rho_{\mathrm{crit,0}} = \Omega_{\mathrm{m, 0}}(1+z)^3$. In a flat universe the local matter and dark energy components are related by $\Omega_{\mathrm{m, 0}} + \Omega_{\mathrm{DE, 0}} = 1$. Regarding the reference $\Lambda$CDM-model, the Friedmann-Robertson-Walker (FRW) equation is then given by
      \begin{equation}
						E^2(z)
   		     = (1 - \Omega_{\Lambda, 0}) (1+z)^3
   		     + \Omega_{\Lambda,0}
   		  \ \ ,
   		\label{frw1Lambda}                	
   		\end{equation}  
where $E(z) \equiv H(z)/H_0$ denotes the expansion rate normalized to its local value.
The respective FRW-equation of the effective vacuum model can be written as 
      \begin{equation}
   		      E^2(z)  
   		      = \frac{(1 - \Omega_{\mathrm{vac,0}})}{1-\Omega_{\mathrm{vac,0}} \, W(E(z))/W_0} \,(1+z)^3\ \ ,
   		\label{frw1vac}                	
   		\end{equation}  
where (\ref{omegavac}) and (\ref{omegavac0}) have been used. 

Note that the explicite value of the local Hubble rate $h_{100}$ does not enter into expression (\ref{frw1vac}): The local amplitude $\Omega_{\mathrm{vac,0}}$ (\ref{omegavac0}) only depends on the relative mass of the lightest neutrino and in general on the structure parameter $\eta$;
the normalized total weighting function $W(z)/W_0$ is just a function of the ratio $H(z)/H_0$ and depends on the relative mass $x_i$ and the degrees of freedom $g_i$ via the sum (\ref{Wdef}).

Regarding the structure parameter $\eta$, let us assume for a moment that even the mass of the lightest neutrino was much greater than the local energy scale $D_0/c^2$. This would imply that for a wide redshift range, $z \leq 1.5$ for instance, only the vacuum field of the photons would contribute, i.e., $W(E) =  W_0 = g_{\mathrm{ph}}$. As has been discussed in section 
\ref{sec:local universe}, we then could still match the resulting effective vacuum energy density $\Omega_{\mathrm{vac,0}}$ with $\Omega_{\Lambda, 0}$ (\ref{OmegaLambda}) assuming that the structure parameter is $\eta > 2.4$. However, if only the photon field was relevant, one would have $W(E(z))/W_0 = 1$ in (\ref{frw1vac}) and therefore $E^2(z) = (1+z)^3$. This is equivalent to a FRW-equation which describes the evolution of a Einstein-de Sitter (EdS) universe. However, the EdS-model is apparently not compatible with current observations [Frieman et al. (\cite{fth2008})]. Within the frame of the present model, this contradiction indicates that the mass of the lightest neutrino is actually below the scale $D_0/c^2$ and that the structure parameter is $\eta <2.4$. This finding supports our presumption made in section \ref{sec:local universe}, namely that $\eta = 1$. Thus, in the remainder of this paper we confine ourselves to considering this constant value.

As shown below, the (normalized) expansion rate $E(z)$, given by expression (\ref{frw1vac}), increases with growing redshift $z$. This implies that the characteristic energy scale $D(E)$ (\ref{DHevolv}) increases as well and that the relative mass of each elementary particle $x_{\mathrm{i}} \propto D^{-1}$ becomes smaller. So, the higher the redshift the more vacuum fields gradually become gravitationally effective and therefore have to be taken into account in the sum $W(E)$ (\ref{Wdef}). 
Especially, if the expansion rate $H(z)$ is so high that all three neutrino mass eigenstates contribute to the effective vacuum, the sum $W(E)$ amounts to 
   \begin{equation}
   		  W(E) =  g_{\mathrm{ph}} - \left|g_{\nu}\right|	
   		  \left(f_{\nu, \,\mathrm{l}}\, (E)		
   		  + f_{\nu, \,\mathrm{2}}\, (E)	
   		  + f_{\nu, \,\mathrm{3}}\, (E)
   		  \right)			
					,
   		  \label{Wz23}                	
   \end{equation}
where we again adopt $g_{\mathrm{ph}} = 2$ and $\left|g_{\nu}\right| = 2$.

The precise redshift values at which the second and third neutrino fields successively become relevant in (\ref{Wz23}), i.e., where the respective specific weighting factors $f_i$ become non-zero, depend on the ordering of the neutrino mass eigenstates. Yet, the neutrino mass ordering is still an open question [e.g. Bilenky (\cite{bilenky2012})]. So we have to consider in the following normal as well as inverted mass ordering. 

		\graphicspath{{d:/research/TeX/PhysRev/PRD/publ/}}
		\begin{figure}
				\includegraphics[width=\linewidth]{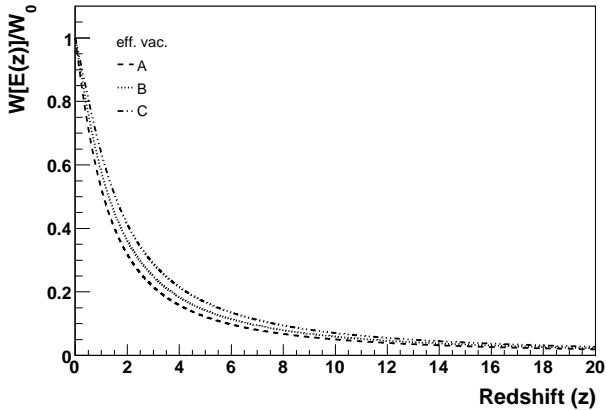}
		 		\caption{Amplitude of the normalized total weighting function $W(E(z))/W_0$ as a function of redshift. In models A - C inverted neutrino mass ordering has been adopted (see table \ref{table:1}).}
		\label{fig:47WE}
		\end{figure}

 		\begin{table}
		\caption{Effective vacuum models. In all models the structure parameter (\ref{DeltaL}) is chosen to be $\eta = 1$.}
 		\begin{ruledtabular}
		 			\begin{tabular}{c c c c c c c c}
				 Model & $x_{\nu, \,\mathrm{l,0}}$  & Mass Order &  $g_{\nu}$		&   $\Omega_\mathrm{vac, 0}$ &  $w_{\mathrm{0}}$ 
				 &  $w_{\mathrm{A}}$ & $H_0 T_{\mathrm{age}}(0)$\\
				 A 	& 	0.476		&		inv.		&		-2		&		0.725		&		-0.93	&		0.13	&	0.933\\
				 B 	& 	0.500		&		inv.		&		-2		&		0.803		&		-0.94	&		0.12	&	1.015\\
				 C 	& 	0.518		&		inv.		&		-2		&		0.864		&		-0.95	&		0.11	&	1.109\\
				 D 	& 	0.476		&		normal		&		-2		&		0.725		&		-0.93	&		0.13	&	1.086\\
				 E 	& 	0.500		&		normal		&		-2		&		0.803		&		-0.94	&		0.12	&	1.167\\
				 F 	& 	0.518		&		normal		&		-2		&		0.864		&		-0.95	&		0.11	&	1.262\\
				 \end{tabular}
 		\end{ruledtabular}
 		\label{table:1}
 		\end{table}

\begin{figure}
				\includegraphics[width=\linewidth]{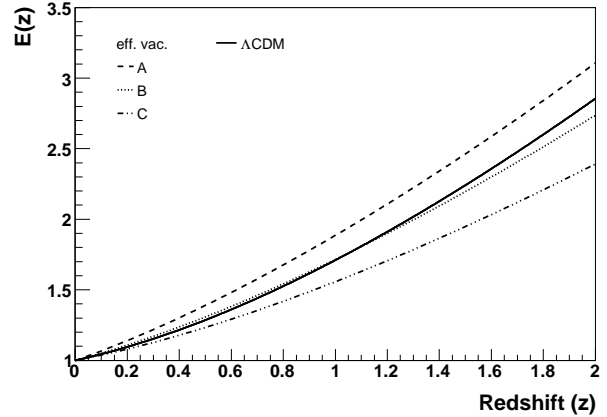}
		 		\caption{Normalized expansion rate $E(z) = H(z)/H_0$ as a function of redshift in the effective vacuum models A - C (inverted mass ordering; cf. table \ref{table:1}) and in the $\Lambda$CDM model (solid line), for comparison.}
		\label{fig:46H}
		\end{figure}

\begin{figure}
				\includegraphics[width=\linewidth]{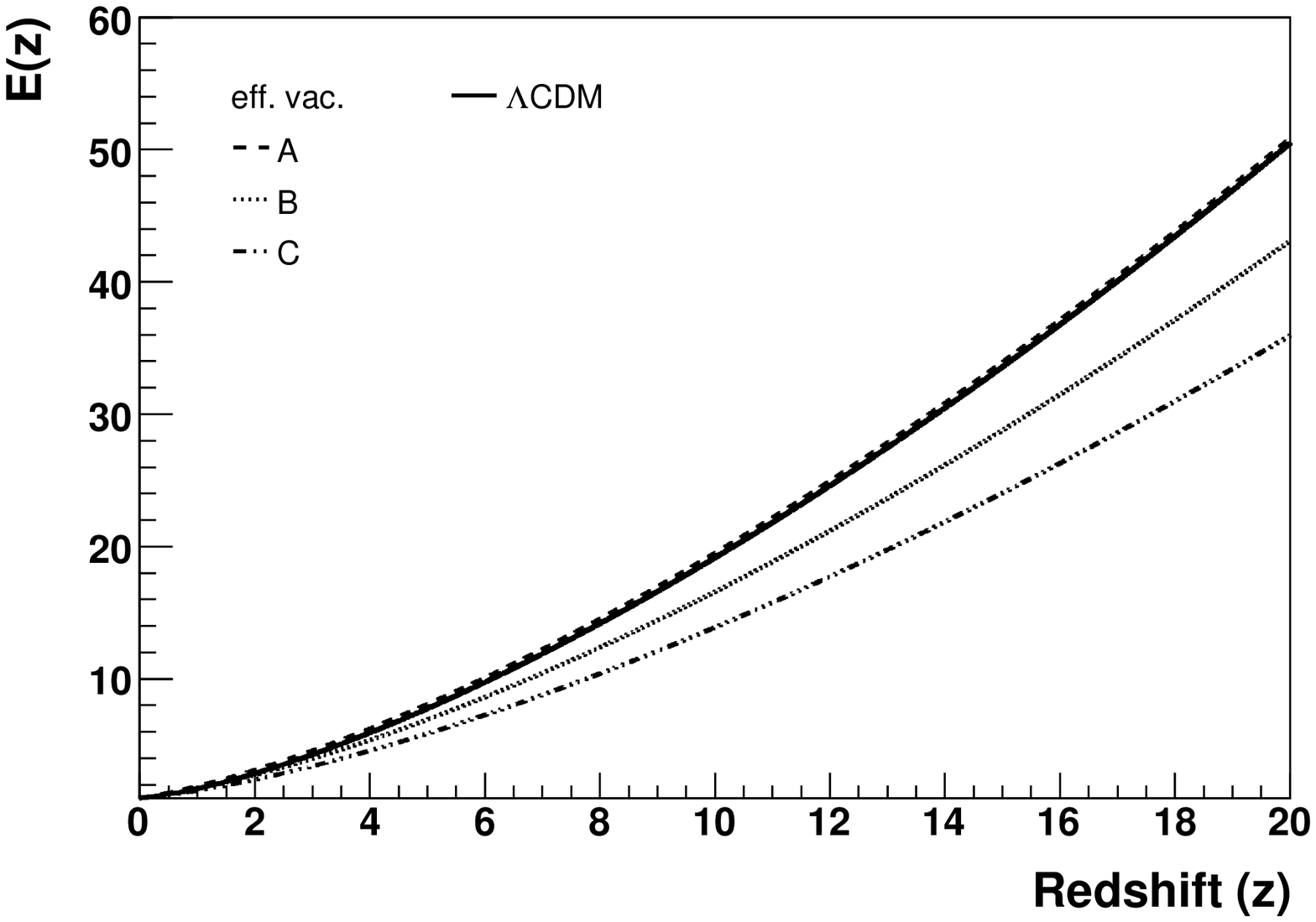}
		 		\caption{Normalized expansion rate $E(z) = H(z)/H_0$ as a function of redshift in the effective vacuum models A - C (inverted mass ordering; cf. table \ref{table:1}) and in the $\Lambda$CDM model (solid line), for comparison. Same as figure \ref{fig:46H} but for an extended redshift range.}
		\label{fig:47H}
		\end{figure}

\begin{figure}
				\includegraphics[width=\linewidth]{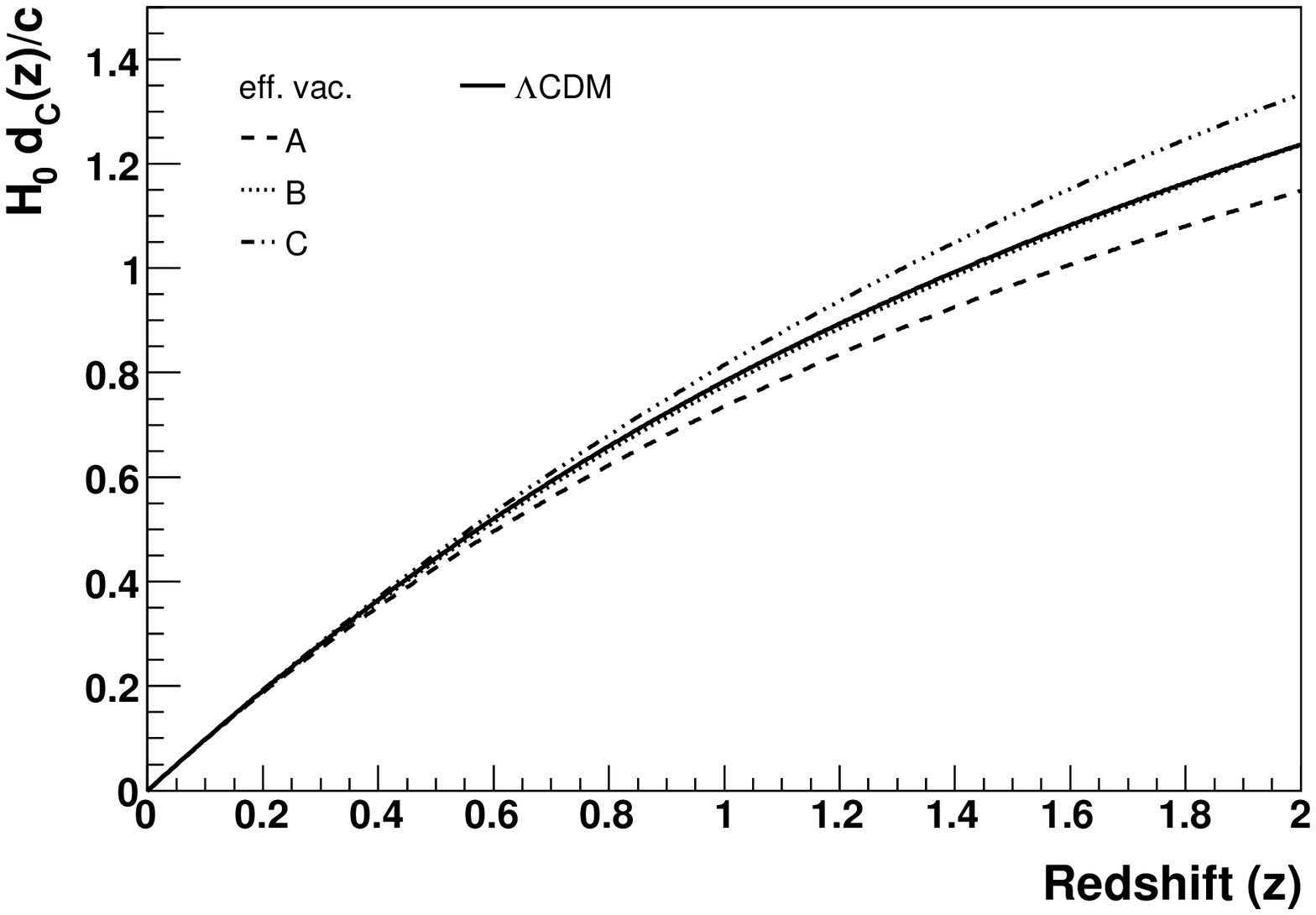}
		 		\caption{Normalized comoving distance (\ref{dCdef}) as a function of redshift in the effective vacuum models A - C (inverted mass ordering; cf. table \ref{table:1}) and in the $\Lambda$CDM model (solid line), for comparison.}
		\label{fig:46r}
		\end{figure}

\begin{figure}
				\includegraphics[width=\linewidth]{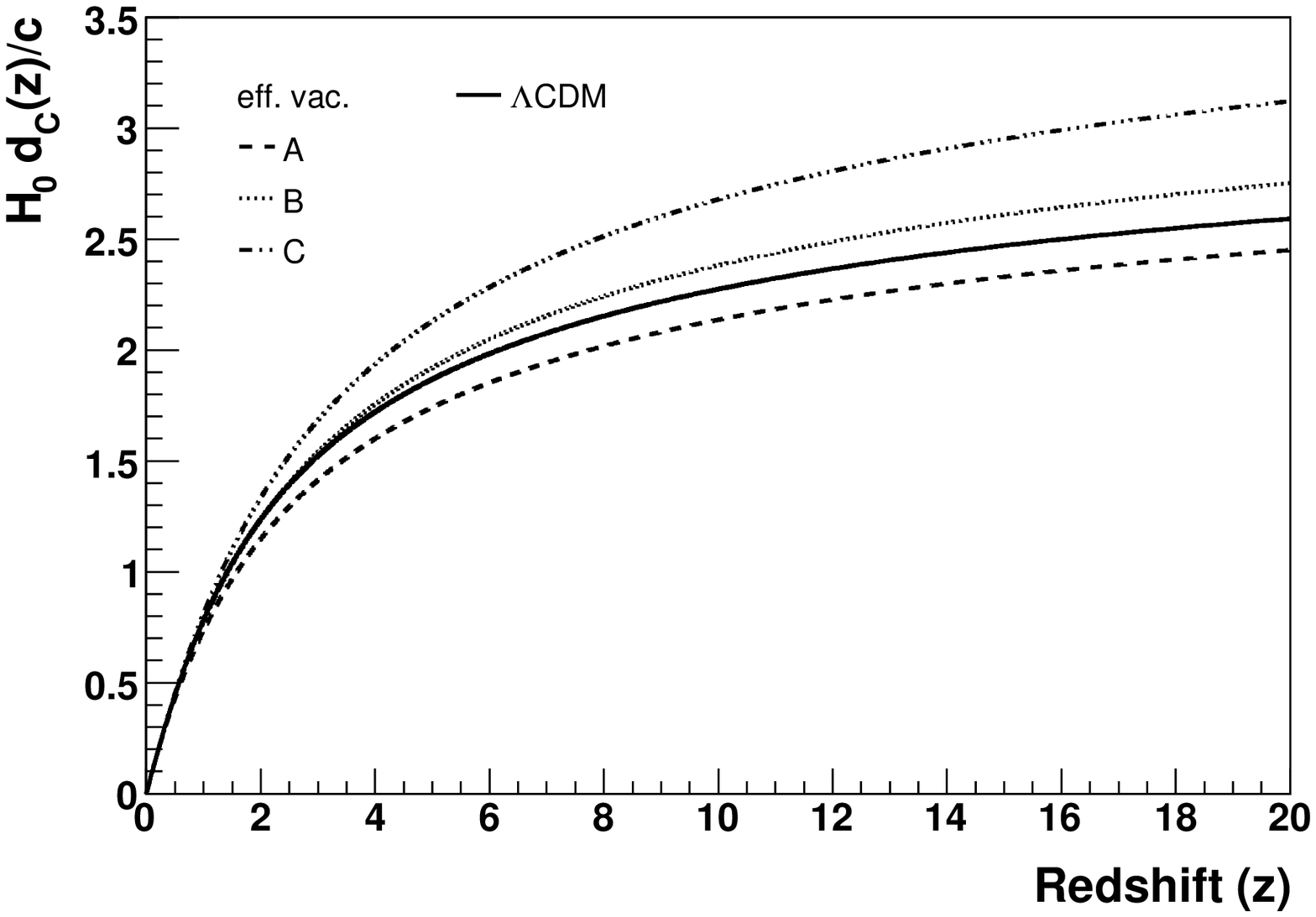}
		 		\caption{Normalized comoving distance (\ref{dCdef}) as a function of redshift in the effective vacuum models A - C (inverted mass ordering; cf. table \ref{table:1}) and in the $\Lambda$CDM model (solid line), for comparison. Same as figure \ref{fig:46r} but for an extended redshift range.}
		\label{fig:47r}
		\end{figure}

In the frame of the present model, inverted mass ordering appears to be particularly interesting: The masses of the second and third neutrino fields are then at least $(\left|\Delta m_{A}^2\right|)^{1/2} \approx 0.05$ eV. Comparing with $D_0$ (\ref{DH0}) and considering (\ref{DHevolv}) and (\ref{xdef}) this implies that the two heavier neutrino mass eigenstates first come into play, if the expansion rate is as large as $H \gtrsim 80 \, H_0$. Below that value, only the vacuum fields of the photon and of the lightest neutrino prevail, as is the case in the local universe. This has an important consequence: Starting from its local value, the neutrino weighting factor $f_{\nu, \,\mathrm{l}}\, (E)$ in (\ref{Wz23}) gradually  approaches unity for increasing $E$ ensuing that the ratio $W/W_0$ becomes quite small at high $z$ (see figure \ref{fig:47WE}). Thus, for sufficiently high redshift, in both the $\Lambda$CDM and the effective vacuum model the relative expansion rates $E(z)$ [cf. eqs. (\ref{frw1Lambda}) and (\ref{frw1vac})] are then predominantly governed by the evolution of the mass density. 
Figures \ref{fig:46H} and \ref{fig:47H} display the normalized expansion rate versus redshift for three effective vacuum models A - C (see table \ref{table:1}) and the $\Lambda$CDM-model for comparison. Figures \ref{fig:46r} and \ref{fig:47r} depict the respective normalized comoving distance $H_0 \,d_{\mathrm{C}}(z)/c$, where $d_{\mathrm{C}}(z)$ is defined by
      \begin{equation}
   		      d_{\mathrm{C}}(z) = \frac{c}{H_0} \int_0^z \frac{\mathrm{d}z\,'}{E(z\,')} \
   		     .
   		\label{dCdef}                	
   		\end{equation} 
In model A we constrained the lightest relative neutrino mass $x_{\nu, \mathrm{l, 0}}$ such that the amplitude of the local effective vacuum energy density $\Omega_{\mathrm{vac,0}}$ matches with (\ref{OmegaLambda}). However, compared to the $\Lambda$CDM-model, one would then find a slightly too strong expansion rate within the redshift range $z \leq 2$. But at even higher redshifts (figure \ref{fig:47H}), the expansion rates of both models behave more or less identically. Due to the enhanced expansion rate, the comoving distance in model A is slightly smaller than that given by the $\Lambda$CDM-model (figsures \ref{fig:46r} and \ref{fig:47r}). In model B we choose a value of the relative neutrino mass such that within the observationally important redshift range $z \leq 2$ the evolution of the comoving distance (figure \ref{fig:46r}) becomes nearly indistinguishable from that of the reference model. However, one would then derive a local effective vacuum energy density of $\Omega_{\mathrm{vac,0}} = 0.80$. Model C demonstrates how sensitively the effective vacuum model reacts on even smallest changes of the relative neutrino mass.

		\graphicspath{{d:/research/TeX/PhysRev/PRD/publ/}}

		\graphicspath{{d:/research/TeX/PhysRev/PRD/publ/}}

		\graphicspath{{d:/research/TeX/PhysRev/PRD/publ/}}

		\graphicspath{{d:/research/TeX/PhysRev/PRD/publ/}}

		\graphicspath{{d:/research/TeX/PhysRev/PRD/publ/}}
		\begin{figure}
				\includegraphics[width=\linewidth]{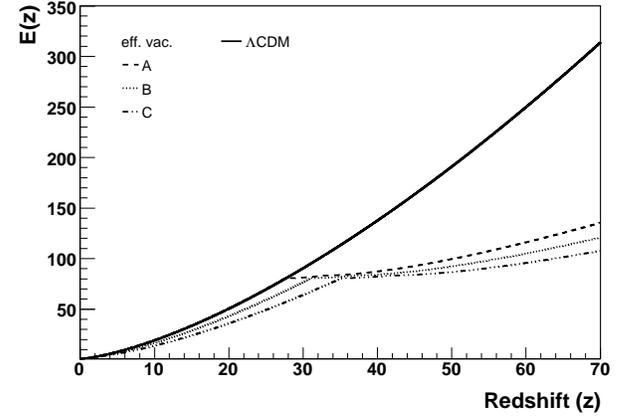}
		 		\caption{Normalized expansion rate $E(z) = H(z)/H_0$ as a function of redshift in the effective vacuum models A - C (inverted mass ordering; cf. table \ref{table:1}) and in the $\Lambda$CDM model (solid line), for comparison. Same as figure \ref{fig:47H} but for an extended redshift range.}
		\label{fig:48H}
		\end{figure}

If the expansion rate is so high that the two other neutrino mass eigenstates come into play, the expansion history deviates strongly from that in a $\Lambda$CDM-model (figure \ref{fig:48H}). The value of the total weighting function $W(E)$ (\ref{Wz23}) and hence the effective vacuum energy density (\ref{rhovactotcrit}) then become negative. This ensues a considerably smaller expansion rate compared to that in the $\Lambda$CDM-model. Nevertheless, even in the effective vacuum model the expansion rate increases with rising redshift. As long as one only considers photons and neutrinos, the normalized total weighting function $W(E)/W_0$ in the denominator of the rhs of expression (\ref{frw1vac}) eventually approaches the constant value of $(g_{\mathrm{ph}} - 3 \left|g_{\nu}\right|/W_0)$, which is about $-10$. This involves that at very high redshift, $z \gtrsim 60$, the expansion rate still develops proportionally to $(1+z)^{3/2}$, but in a way as if the mass density was effectively reduced by a certain factor. Compared to the findings in a $\Lambda$CDM-model the resulting expansion rate is then diminished by a factor of approximately $1/\sqrt{1+10\Omega_{\mathrm{vac,0}}} \approx 0.35$.

Of course, this inferred reduction factor of the expansion rate is based on our simplifying assumption that besides photons and neutrinos no other elementary particle fields, e.g. bosonic axions, become relevant according to (\ref{xdef}). The latter would happen if the mass of those particles were smaller than the characteristic cosmic energy scale $D(H)$ at a given expansion rate. The expansion rate at the era of recombination is of special interest. The corresponding redshift is ${z_\mathrm{rec}} \approx 1100$, which involves $E({z_\mathrm{rec}}) \approx 6700$ if only photons and neutrinos are taken into account. From (\ref{DHevolv}) one then obtains $D({z_\mathrm{rec}}) \approx 0.45$ eV. Thus, if no other elementary particles with masses below that scale actually exist, only the vacuum fields of the photons and neutrinos contribute to the effective vacuum density at least back until the recombination era. However, this finding certainly can not be true for much earlier cosmic times. Otherwise one would infer a too low expansion rate in the early universe which implies that the universe was much older at the time of recombination than in the $\Lambda$CDM-model. This would presumably conflict with the observed fluctuation spectrum of the cosmic microwave background (CMB) \cite{komatsu2011}. So, in the framework of the present model, one would conclude that bosonic particles must exist within the considered low mass range.
In principle, the CMB can offer important clues to this issue; a further consideration of this question is, however, beyond the scope of the present paper.

The graphs in figure \ref{fig:48H} exhibit a peculiar behaviour: Whenever the expansion rate is such that the energy scale $D(E)$ (\ref{DHevolv}) matches the mass energy of a specific neutrino mass eigenstate, the evolution of the expansion rate exhibits a sudden change. However, this sudden change of the gradient is the consequence of just the mathematically lowest-order treatment of the proposed model. This rather unphysical effect would vanish in a more thorough analysis, where one should account for, e.g., the probabilistic nature of the lifetime of the virtual particles leading to a softening of the sharp cutoff of the specific weighting function $f_{\mathrm{i}}$ at $x_{\mathrm{i}} = 1$ (see figure \ref{fig:fi}). Yet, the general structure of the function $E = E(z)$ would persist.   

\begin{figure}
				\includegraphics[width=\linewidth]{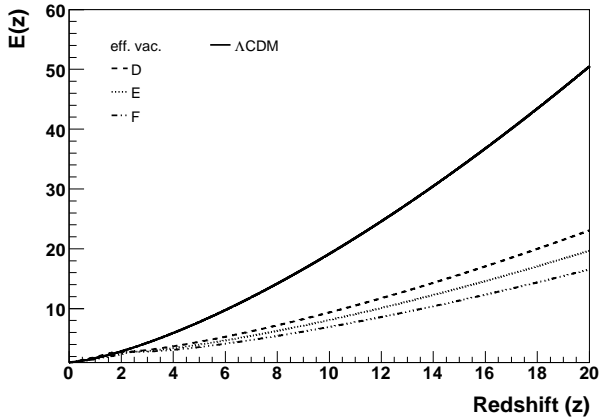}
		 		\caption{Normalized expansion rate $E(z) = H(z)/H_0$ as a function of redshift in the effective vacuum models D - F (normal mass ordering; cf. table \ref{table:1}) and in the $\Lambda$CDM model (solid line), for comparison.}
		\label{fig:50H}
		\end{figure}

So far, we considered inverted neutrino mass ordering. Figure \ref{fig:50H} displays the resulting development of the expansion rate for normal mass ordering (models D - F) where we adopted, except for the mass ordering, the same model parameters as in figure \ref{fig:48H}. It clearly shows that the deviation to smaller expansion rates as discribed above now already occurs at much smaller redshifts ($z \approx 2)$. The reason for this lies in the fact that the mass of the second massive neutrino is now close to that of the lightest one.

		\graphicspath{{d:/research/TeX/PhysRev/PRD/publ/}}

		\graphicspath{{d:/research/TeX/PhysRev/PRD/publ/}}
		\begin{figure}
				\includegraphics[width=\linewidth]{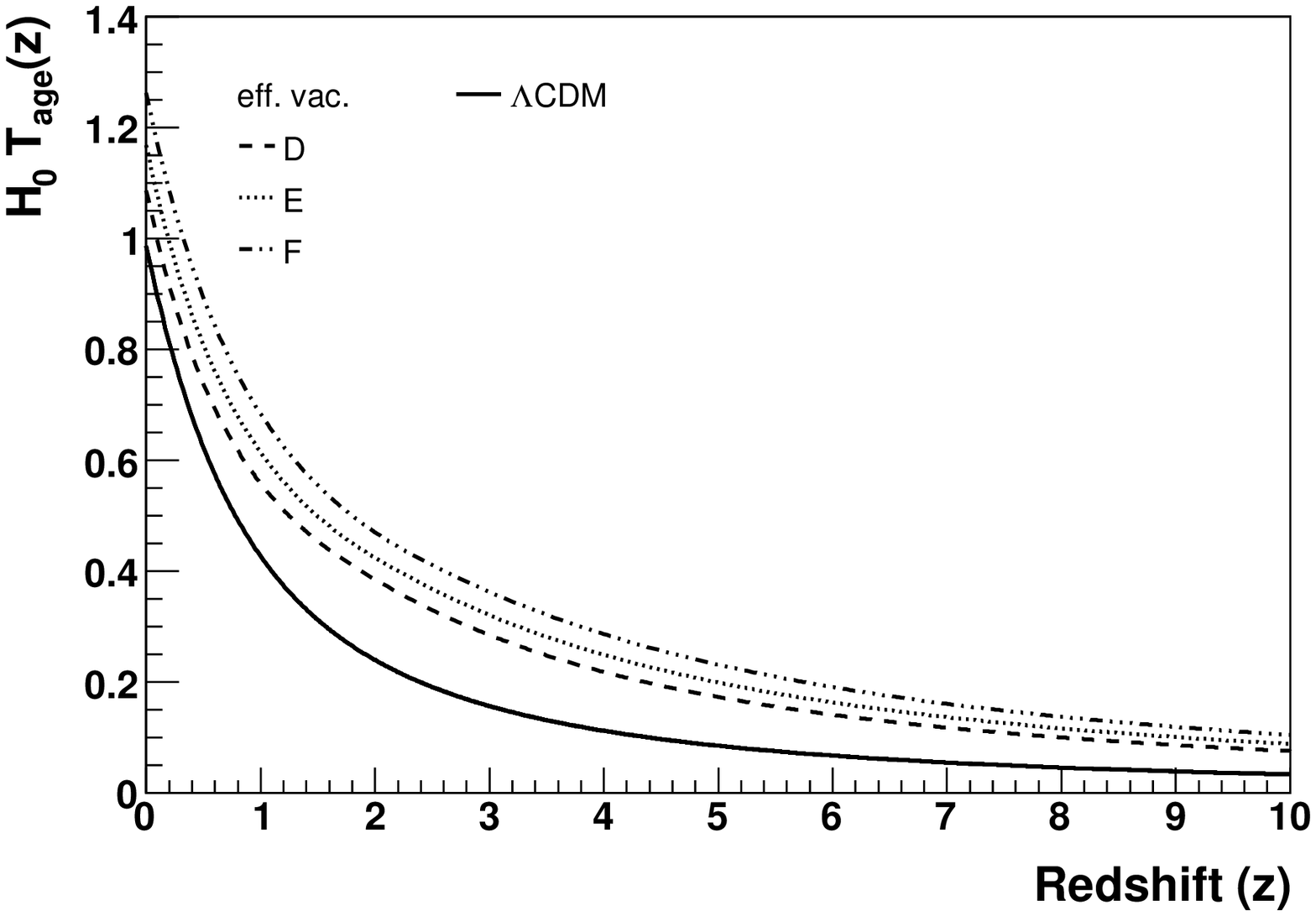}
		 		\caption{Normalized age of the universe (\ref{Tagedef}) as a function of redshift in the effective vacuum models D - F (normal mass ordering; cf. table \ref{table:1}) and in the $\Lambda$CDM model (solid line), for comparison.}
		\label{fig:50T}
		\end{figure}

		\graphicspath{{d:/research/TeX/PhysRev/PRD/publ/}}
		\begin{figure}
				\includegraphics[width=\linewidth]{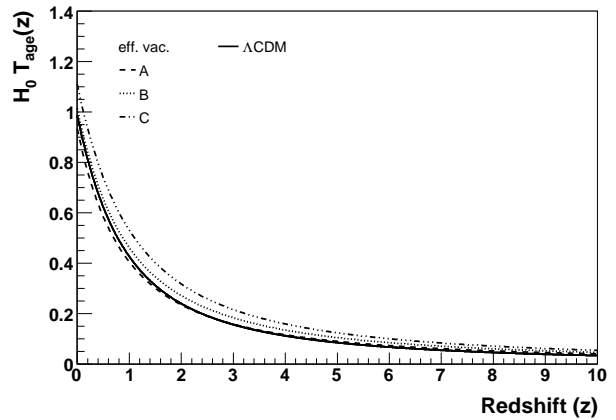}
		 		\caption{Normalized age of the universe (\ref{Tagedef}) as a function of redshift in the effective vacuum models A - C (inverted mass ordering; cf. table \ref{table:1}) and in the $\Lambda$CDM model (solid line), for comparison.}
		\label{fig:48T}
		\end{figure}

Hence, in the case of normal mass ordering the expansion rate is found to be much smaller than the respective outcome in the $\Lambda$CDM-model already within the redshift range $z \lesssim 10$ . This in turn considerably enlarges the inferred age of the universe, $T_{\mathrm{age}}(z)$. For $z \lesssim 10$ a reliable approximation of $T_{\mathrm{age}}(z)$ is given by
       \begin{equation}
   		      T_{\mathrm{age}}(z) \approx \frac{1}{H_0} \int_z^{z_\mathrm{rec}} \frac{\mathrm{d}z\,'}{(1+z\,')E(z\,')} \
   		     ,
   		\label{Tagedef}                	
   		\end{equation}  
where we assume that up to redshift $z_\mathrm{rec}$ the effective vacuum energy density is governed only by the vacuum fields of the photons and the neutrinos. The inferred normalized present day age of the universe, $H_0 T_{\mathrm{age}}(z=0)$, in the effective vacuum models is given in the last column of table \ref{table:1}. The respective value in the $\Lambda$CDM-model is 0.988.

Figure \ref{fig:50T} displays the normalized age as a function of redshift in models D - F. One then derives that at $z = 4$ the age was about twice as high as in the $\Lambda$CDM-model. This could help to alleviate the cosmic age problem from which the $\Lambda$CDM-model is supposed to suffer, as it apparently fails to reconcile the age of an old quasar \cite{yang2010}, \cite{wang2010}. In this respect, models D - F (normal mass ordering) might be more advantageous than models A - C (inverted mass ordering). The evolution of the respective normalized age in the latter models is shown in figure \ref{fig:48T}, for comparison. On the other hand, a too low expansion rate in the range $z \lesssim 10$ could conflict with the usually adopted successful concept of cosmic structure formation. A further consideration of this important question, however, needs a more thorough treatment and has to be deferred to later work (see also section \ref{sec:inhomogeneity}).

Of course, the current unknowns of elementary particle physics enter into the present model as (still) free parameters. Yet, the considerations of this section demonstrate that the gravitationally effective vacuum model of dark energy allows the construction of a consistent model that describes a cosmic expansion history which is largly identical to the one given by the $\Lambda$CDM-model. That means that the proposed model can be regarded to be, at least in principle, in accord with current observational data.

\section{Energy Conservation} 
\label{sec:Econserv}

In this section we consider the energy conservation of the effective vacuum component and its equation-of-state (EOS) parameter, $w \equiv P/\rho$, where $P$ denotes the pressure.

As is well-known, a non-zero cosmological constant $\Lambda$ may be regarded as the action of a constant dark energy density with a specific EOS parameter of 
      \begin{equation}
					w_{\Lambda} = - 1
					. 
			\label{wLambda}
			\end{equation}
At present, the observational data are consistent with a constant $w$-parameter of this value [e.g. \cite{komatsu2011}],
hence apparently supporting the $\Lambda$CDM-model.

It is commonly suggested that a value of $w = - 1$ characterizes the (constant) vacuum state. This relies on the ansatz $d(\rho_{\mathrm{vac}} a^3) + p_{\mathrm{vac}} d(a^3) = 0$ describing energy conservation of the vacuum `fluid' in an expanding volume. In this respect it becomes clear why the cosmological constant, $\Lambda$, is usually associated with the otherwise unspecified action of the vacuum. From that point of view, it is tempting to consider the proposed \emph{effective} vacuum model just as another variant of a specific class of $\Lambda$CDM-models, namely of those models that allow for a time varying $\Lambda(t)$. However, as we would like to stress, the effective vacuum model is essentially different from all $\Lambda$CDM-models. Indeed, its underlying concepts make it different to most of the cosmological models proposed in the literature as we try to expound in the following. 

The problem centres on the issue of energy conservation. One commonly begins with the ansatz
      \begin{equation}
				\mathrm{d} \left(\rho_{\mathrm{tot}} \, a^3 \right)/{\mathrm{d t}}
					+ P_{\mathrm{tot}} \, {\mathrm{d}}a^3 /{\mathrm{d t}} 
					= 0
					,
			\label{Etotconserv}
			\end{equation}
where $\rho_{\mathrm{tot}}$ and $P_{\mathrm{tot}}$ denote the total energy density and pressure of all cosmic constituents like matter, radiation and dark energy. For the sake of simplicity, we assume again
      \begin{equation}
					\rho_{\mathrm{tot}} = \rho_{\mathrm{m}} + \rho_{\mathrm{DE}},
					\ \ \ \ \ \ \mathrm{and}\  \ \ \ \ \
					P_{\mathrm{tot}} = P_{\mathrm{m}} + P_{\mathrm{DE}}
										.
			\label{rhotot}
			\end{equation}
From (\ref{Etotconserv}) and (\ref{rhotot}) one finds, in general,
      \begin{equation}
					{\mathrm{d}} \left(\rho_{\mathrm{m}} a^3 \right)/{\mathrm{d t}}
					+ P_{\mathrm{m}}\; {\mathrm{d}} \left(a^3 \right)/{\mathrm{d t}}
					= - \dot{S}
					\ ,
			\label{Mconserv}
			\end{equation}
and 
      \begin{equation}
					{\mathrm{d}} \left(\rho_{\mathrm{DE}} a^3 \right)/{\mathrm{d t}} 
					+ P_{\mathrm{DE}}\; {\mathrm{d}} \left(a^3 \right)/{\mathrm{d t}}
					=  \dot{S}
					\ .
			\label{DEconserv}
			\end{equation}
Note that for non-relativistic matter (luminous and dark), the pressure term in (\ref{Mconserv}) can be neglected. 			
In (\ref{Mconserv}) and (\ref{DEconserv}), $\dot{S}$ represents a possible net energy exchange rate between the matter and the dark energy components which cannot be excluded from the outset. Employing the definition $w_{\mathrm{DE}} \equiv P_{\mathrm{DE}}/\rho_{\mathrm{DE}}$, the latter equation can be written as
      \begin{equation}
   		   \dot{\rho}_{\mathrm{DE}} + 3H \rho_{\mathrm{DE}} ( 1 + w_{\mathrm{DE}}) 
   		   = \dot{S} a^{-3}
   		   ,
   \label{DEdot}                	
   \end{equation} 
where the over-dot denotes the derivative with respect to time. 
 
If one demands energy conservation for each component separately, i.e., $\dot{S} = 0$, the evolution of the dark energy density (\ref{DEdot}) is completely governed by the EOS parameter, $w_{\mathrm{DE}}$, which, in general, can itself develop in the course of the cosmic evolution. Such an ansatz is employed, for instance, in quintessence models 
[e.g. \cite{calkam2009}, and references therein]. 
In the case of the $\Lambda$CDM-model, where $w_{\mathrm{DE}} = w_{\Lambda} = - 1$ is a constant, one yields the aforementioned constancy of the dark energy density. 

On the other hand, in time varying $\Lambda(t)$CDM-models one still puts $w_{\mathrm{DE}} = - 1$ in eq. (\ref{DEdot}). But in order to accomplish the modelling of an evolving dark energy density one has to allow for an, otherwise unspecified, energy exchange with the matter component; that means, one puts $\dot{S}(a) \neq 0$ on the rhs of (\ref{Mconserv}) and (\ref{DEdot}) 
[e.g. \cite{basilakos2009}].
     
None of these procedures appear to be viable for the effective vacuum model: An energy exchange between the effective vacuum and the matter component does not make sense, since the vacuum fluctuations are supposed to be affected only by the expansion of space but not by couplings to some other fields. This would involve $\dot{S} = 0$ and $w_{\mathrm{DE}} \neq - 1$ in (\ref{DEdot}). Though one could formally derive a $w$-parameter function
      \begin{equation}
 					w_{\mathrm{DE}}(z) = - 1 - \dot{\rho}_{\mathrm{DE}}(z)/(3H \rho_{\mathrm{DE}}(z)), 
   \label{wDEdef}                	
   \end{equation} 
one would be then faced with the unphysical result that $w_{\mathrm{DE}}(z)$ diverges to infinity as $\rho_{\mathrm{vac}}(z)$ (\ref{rhovactotcrit}) becomes zero at some particular redshift. This is not a mere technical problem but a consequence of the underlying conceptual inconsistency.

However, this inconsistency resolves when recalling that the effective vacuum energy, becoming noticible as dark energy, is only part of the enormous reservoir of the total vacuum with which the effective component exchanges energy. To be more specific: The effective vacuum energy arises from the virtual vacuum by virtue of the presumed interaction with expanding space. In a sense, the energy content of the effective vacuum can be regarded as though it is generated from `nothing'; so there is no need for energy exchange with any other component, like matter, radiation and so forth . This also implies that the total net rate at which the amplitude of the effective vacuum component may change in time is basically defined by $\dot{\rho}_{\mathrm{vac}}$ itself. 

Thus, we propose the following point of view: Whilst the matter component is a closed, energy conserving subsystem, i.e., $\dot{S} = 0$ in (\ref{Mconserv}), the effective vacuum component couples with the virtual reservoir via a net energy exchange rate $\dot{S}_{\mathrm{vac}}$. As pointed out above, this rate is necessarily given by $\dot{S}_{\mathrm{vac}} = \dot{\rho}_{\mathrm{vac}} \, a^3$. Hence, eq. (\ref{DEdot}) can be written as
\begin{equation}
   		   \dot{\rho}_{\mathrm{vac}} + 3H \rho_{\mathrm{vac}} ( 1 + w_{\mathrm{vac}}) 
   		   = \dot{S}_{\mathrm{vac}} \,  a^{-3} = \dot{\rho}_{\mathrm{vac}}
   		   .
   \label{vacdot}                	
   \end{equation} 
From (\ref{vacdot}), we deduce a \emph{constant} EOS parameter of 
      \begin{equation}
					w_{\mathrm{vac}} = - 1
					. 
			\label{wvac}
			\end{equation}

For reasons of energy conservation, one could formally supplement eq. (\ref{vacdot}) by 
\begin{equation}
   		   \dot{\rho}_{\mathrm{vir}} + 3H \rho_{\mathrm{vir}} ( 1 + w_{\mathrm{vir}}) 
   		   = - \dot{\rho}_{\mathrm{vac}} = \dot{\rho}_{\mathrm{vir}}
   		   ,
   \label{vacvirdot}                	
   \end{equation} 
where ${\rho}_{\mathrm{vir}}$ denotes the remaining virtual part of the total vacuum energy density. The latter relation in (\ref{vacvirdot}) follows from the presumption that the \emph{total} vacuum energy density is a constant. Again, one would derive a constant EOS parameter of $w_{\mathrm{vir}} = - 1$. The sum of eqs. (\ref{vacdot}) and (\ref{vacvirdot}) would then describe conservation of energy of the complete vacuum sector. 
Yet, in the real physical world only the \emph{effective} vacuum component (\ref{vacdot}) becomes noticable as dark energy, whilst the \emph{virtual} component (\ref{vacvirdot}) does literally not appear. Hence, in reality one necessarily deals with only a subsystem. This has the important consequence that the universe, as it is usually considered to comprise of matter, radiation, dark energy and so forth, is essentially a (thermodynamically) open system which exchanges energy (and as well momentum) with the immense reservoir of the virtual vacuum. 

This becomes more clear considering the time derivative of the FRW equation. If one assumes adiabaticity (cf. eq. \ref{Etotconserv}), one usually arrives at
      \begin{equation}
   		    \frac{\ddot{a}}{a} =
   		    \frac{8 \pi G}{3 c^2}
   		     \left[ -\frac{1}{2} \rho_{\mathrm{m}}(z) 
   		     - \frac{1}{2} \rho_{\mathrm{DE}}(z) \,(1 + 3 w_{\mathrm{DE}})
   		     \right]
   		   .
   		\label{frw2DE}                	
   		\end{equation}
But adopting (\ref{vacdot}), characterized by its non-vanishing source term, we obtain 
      \begin{equation}
   		    \frac{\ddot{a}}{a} =
   		    \frac{8 \pi G}{3 c^2}
   		     \left[ -\frac{1}{2} \rho_{\mathrm{m}}(z) 
   		     + \rho_{\mathrm{vac}}(z) 
   		     +  \frac{1}{2H} \dot{\rho}_{\mathrm{vac}}(z)
   		     \right]
   		   ,
   		\label{frw2}                	
   		\end{equation}
where (\ref{wvac}) has been employed.  The last term in the brackets stems from the coupling with the virtual vacuum. It is this dissipative term that makes the effective vacuum model different to not only the $\Lambda(t)$CDM-model but to basically all adiabatic models. It vanishes in the special case of constant expansion rate, since the amplitude of the effective vacuum energy density is then a constant. In the local universe, where the expansion rate changes slowly, the value of this extra energy contribution is indeed non-zero but much smaller than the amplitudes of the other energy components. However, in a situation where the cosmic expansion rate evolves on short time-scales, all energy contributions could be of the same order of magnitude; this might have been the case, for instance, in the early universe.

Note that due to the thermodynamic openness, the amplitude of the effective vacuum energy density $\rho_{\mathrm{vac}}(z)$ evolves in the course of the cosmic expansion history, though its EOS parameter is $w_{\mathrm{vac}} = - 1$. From an observational point of view, the evolution of $\rho_{\mathrm{vac}}(z)$ can be regarded to mimic an effective $w$-parameter $w_{\mathrm{DE}}(z) \neq -1$ given by (\ref{wDEdef}). However, as mentioned above, such a derived $w_{\mathrm{DE}}(z)$ would not behave physically meaningfully at any redshift value. Yet, regarding the local universe, it allows to test the expected outcome of the proposed model by comparing it with observational findings. 
In observational cosmology, the most commonly used phenomenological description of dark energy consists in a linear parameterization of \cite{chevallier2001}, \cite{linder2003} 
      \begin{equation}
   		     w(a) = w_0 + w_{\mathrm{A}} (1 - a/a_0)
   		   ,
   		\label{wparameter}                	
   		\end{equation}
where $w_0$ and $w_{\mathrm{A}}$ are constants. 
In order to evaluate these quantities in terms of the effective vacuum model, we apply an expansion of $w_{\mathrm{DE}}(a)$ about $a = a_0$ up to first order. That means, we write
      \begin{equation}
   		     w_{\mathrm{DE}}(a) \approx w_{\mathrm{DE}}(a_0) + \left. \frac{\mathrm{d}w_{\mathrm{DE}}(a)}{\mathrm{d}(1-a/a_0)} \right|_{a=a_0} (1 - a/a_0)
   		   .
   		\label{wexpand}                	
   		\end{equation}
Identifiying the expansion parameters in (\ref{wexpand}) with the respective quantities in (\ref{wparameter}) and employing (\ref{wDEdef}), one obtains the relations
      \begin{equation}
   		     w_0 = - 1 +(1/3) \,\rho'_{\mathrm{DE}}(z=0)/\rho_{\mathrm{DE,0}}
   		\label{w0def}                	
   		\end{equation}
and
      \begin{equation}
   		     w_{\mathrm{A}} = - 3(1+w_0)^2 + w_0 +1 + (1/3)\, \rho''_{\mathrm{DE}}(z=0)/\rho_{\mathrm{DE,0}} \;
   		   ,
   		\label{wAdef}                	
   		\end{equation}
where the prime denotes the derivative with respect to $z$. Finally, we numerically determine the derivatives of $\rho_{\mathrm{vac}}(z)$ at $z=0$ and insert these numbers into (\ref{w0def}) and (\ref{wAdef}). The resulting values of $w_0$ and $w_{\mathrm{A}}$ are given in table \ref{table:1}. The numbers show that within the frame of the effective vacuum model one expects $w_0 \gtrsim -1$ and a small (positive) value of $w_{\mathrm{A}}$ of one tenth. This appears to be in good agreement with recent observational determinations which point to values $w_0 = - 0.93 \pm 0.13$ and $w_{\mathrm{A}} = -0.4 \pm 0.7$ 
\cite{komatsu2011}.
Though this agreement should not be over-estimated, it at least demonstrates that also in respect to the $w$-parameter the proposed model is consistent with current observations.		

\section{Inhomogeneities}
\label{sec:inhomogeneity}

So far we considered the evolution of a homogeneously distributed effective vacuum energy density. However, as has already been argued by Caldwell et al. \cite{caldwell1998}, a spatially uniform but time-evolving cosmic energy component is ill-defined and unphysical.

In terms of the present model, we actually expect the amplitude of $\rho_{\mathrm{vac}}(H)$ to fluctuate, since the rate of expansion of space is presumably inhomogeneous due to the lumpiness of the matter component. Within gravitationally stable configurations, like clusters of galaxies, space expansion is supposedly considerably diminished, thus reducing the amount of $\rho_{\mathrm{vac}}(H)$ within these regions.

Yet, another reason for spatial inhomogeneities of the effective vacuum energy density exists, the analysis of which goes far beyond the scope of the present paper: In section \ref{sec:expanding space time} we assumed that the energy and momentum of the virtual particles are affected by the expansion of space. More generally speaking, we assumed that the virtual particles are affected by a specific deviation from Minkowskian space-time. In this sense, we assert that any deviation from Minkowskian space-time generates a non-zero amplitude of the effective vacuum energy and momentum density. This implies that in a spatially inhomogeneous situation, all components of the respective effective vacuum energy-momentum tensor might be non-zero.

In the special case where the deviation from Minkowskian space-time is caused by a potential gradient, $\nabla \phi$, we expect each term of the tensor components to show basically the same structure as that of $\rho_{\mathrm{vac}}(H)$ given in (\ref{rhovacW}), but where $H$ is replaced by the rate $(\nabla \phi)/c$. Of course, these additional components would cause a backreaction on the evolution of space-time and might as well affect the process of cosmic structure formation; and, in respect to the latter, they might also help to alleviate the dark matter problem. A further consideration of this issue is postponed to later work.

\section{Conclusions}
\label{sec:conclusions}

We conclude that the proposed gravitationally effective vacuum model has the potential to solve the (old) cosmological constant problem. The same holds for the coincidence problem: Since the expected effective vacuum energy density (\ref{rhovactotcrit}) is closely related to the cosmic critical density, it is a natural outcome of this model that one expects both to be of the same order of magnitude. 

The model provides important constraints between cosmological parameters and elementary particle parameters. 
It gives a natural explanation for the observed intriguing coincidence between the cosmic dark energy scale and the scale of the neutrino mass. 
Based on the outcomes of the model, we predict that the mass of even the most massive neutrino is $0.05\; \mathrm{eV}/c^2$ at the most. This value is still below the sensitivity of $0.2\; \mathrm{eV}/c^2$ of the KATRIN experiment \cite{wolf2010}. 

The model is based on the assertion that the assumed discreteness of space-time primarily involves a lower limit of the \emph{changes} of length-scales and time-scales, but not of the scales itself. This conceptual shift from a static to a process-related constraint leads to a natural UV-cutoff (\ref{E2leq}). It also has the important consequence that the presumed process-related microstructure of space-time actually affects macroscopic scales and that this happens especially to be the case in low-energy interactions. This finding is in contrast to the common belief that the discreteness of space-time becomes important only when the involved energies approach the Planck energy scale.   

An essential ingredient of the model is the discrimination between an effective and a virtual vacuum component, where the two components exchange energy and momentum. However, in the real physical universe only the effective vacuum becomes noticable as dark energy. This implies that the universe has to be considered as a (thermodynamically) open subsystem that exchanges energy and momentum with the huge reservoir of the virtual vacuum.

\appendix*
\label{sec:appendix}
\section{Weighting function}

\noindent Let $\xi \equiv \hbar c k /D(H)$, $\chi_{i} \equiv \sqrt{x_{i}^2 + \xi^2}$  
and
		\begin{equation}		
			   f(x_{i}) \equiv	f_{i}  
			   = \frac{1}{16 \pi} 
			      \int_{\xi_{\,0, \, i}}^{\xi_{1,\, i}}
   					 \chi_{i} \: \mathrm{d}^3 \xi 
   			 = \frac{1}{4} 
			      \int_{\xi_{\,0, \, i}}^{\xi_{1,\, i}}
   					 \chi_{i} \: \xi^2 \: \mathrm{d}\xi , 
   \label{kintegral}     	
   \end{equation}
where $x_{i}$ is defined in (\ref{xdef}). The integral $f_{i}$ can be expressed analytically by \cite{grarys1981}
   \begin{equation}			
			f_{i}   
			=
   		(1/32)\left[2\xi \chi_{i}^3 - x_{i}^2 \xi \chi_{i} - x_{i}^4 \ln \left(\xi+\chi_{i} \right) \right]_{\xi_{\, 0, \, i}}^{\xi_{\, 1, \, i}} .
   		 \label{intgeneral}  	
   \end{equation}    
Referring to (\ref{kiminmax}) the limits are $\xi_{0/1, i} = 1 \pm \sqrt{1-x_{i}^2}$. For the special case of a massless field, i.e., $x_{i} = 0$, one finds $f(0) = 1$. In general, it is $0 \leq f_{i} \leq 1$.
   

\begin{acknowledgments}
The author would like to thank W. H. Kegel for helpful discussions and J. Reinhardt for reading a first draft of this paper and for fruitful comments. This work was supported by the Physikalischer Verein, Frankfurt a.M.
\end{acknowledgments}


\end{document}